What Constitutes Peer Review of Data – A survey of published peer review guidelines

By Todd Carpenter[1]

orcid.org/0000-0002-8320-0491

The sharing of research data has exploded in the past decade, and a variety of publications and organizations are putting policies in place that require data publication in some form.

Over the past decade, the number of journals that accept data has increased, as have the number and scope of repositories collecting and sharing research data. Prior to 2010, data sharing was quite limited in scholarly publishing. A 2011 study of 500 papers that were published in 2009 from 50 top-ranked research journals showed that only 47 papers (9%) of those reviewed had deposited full primary raw data online. During the intervening years, the pace of data publishing increased rapidly. As another study notes, the number of data sets being shared annually has increased by more than 400% from 2011 to 2015, and this pace will likely continue. A culture of data sharing is developing, and researchers are responding to data sharing requirements, the efficacy of data sharing, and its growing acceptance as a scientific norm in many fields.

The process is driven in part by both funding and publication policies, which have been encouraging data sharing. The number of titles that explicitly require such sharing in some form is also increasing rapidly. In the past few years, PLOS, AGU, SpringerNature, and the American Economic Association, to highlight just a few, has each put forward policies about data sharing.

In addition, data access has been the focus of other efforts, such as the COPDESS Statement of Commitment, which has 43 signatories. A variety of funding agencies, such as the Wellcome Trust, the Gates Foundation, and the Arnold Foundation now include data sharing as part of their funding policies, and a variety of government agencies are covered by the 2013 OSTP memo on increasing access to federally funded research.

A core element of what distinguishes scholarly publishing from trade efforts is the peer review process. As the availability of research data is increasing, it is important to ask how much of this data is peer reviewed. In a 2014 Study of 4,000 researchers by David Nicholas et al, "[i]t was generally agreed that data should be peer reviewed." But what constitutes peer review of research data? What are existing practices related to peer review of research data sets? Since a number of journals specifically focus on the review and publication of data sets, reviewing their policies seems an appropriate place to start in assessing what existing practice looks like in the "real world" of reviewing and publishing data.

The process of peer review of articles varies from title to title, but it usually consists of two stages: an editorial assessment by the journal's editor or editorial team, then an external evaluation by several peer reviewers. This process is well studied and described in the literature. In the first stage of the process, journal editors focus on whether the paper matches the journal's scope and aims, whether it will be of interest to the readership, whether the author followed basic instructions, and whether the paper meets the journal's minimum quality standards for writing and content. David Schultz described in a 2010 paper that journal editors reject between 6 and 60% of papers at this stage. If a paper moves passes this initial test, it is then sent to independent reviewers for consideration. These reviewers usually provide a deeper analysis and critique of the paper, a step that, according to Walker and Rocha da Silva, involves factors such as study design and methodology, soundness of process and results, data clarity, interpretation of results, completeness of the study, novelty and significance, ethical issues, and other journal-specific criteria. Of course, this process isn't without its critics, its faults, its troubles, or its resulting errors.

At the turn of the decade, a movement began to advance publication of research data sets as first class research objects. This has led to a variety of successful initiatives, such as the Joint Data Citation Principles and the FAIR Principles on data management, that aim to increase awareness and improve practice regarding the publication of datasets. Part of the data publication process is the peer review of those research data sets, which was described in a 2011 article by Lawrence, Jones, Matthews, Pepler, and Callaghan. That article put forward a set of criteria that should be applied in publication of data sets, focusing primarily on metadata and technical review of the material to assess its quality.

I was interested to see how those recommendations were being adopted by journals and whether developing practice might be discerned by examining the available data peer review instructions. Peer review policies are often not overly proscriptive, so as to provide reviewers and editors the opportunity to adjust practice based on the content at hand.

Peer review of data is similar to peer review of an article, but it includes a lot more issues that make the process a lot more complicated. First, a reviewer has to deal with the overall complexity of a research data set—these can be large and complex information objects. Oftentimes, the data go through a variety of pre-processing and error-cleansing steps that should be monitored and tracked. Some data sets are constantly changing and being added to over time, so the question must be asked, does every new study based on a given data set need a new review or could an earlier review still apply? To conduct a proper analysis, the methodology of the data collection should be considered, an examination that can go as deep as describing instrument calibration and maintenance. Even after a data set is assembled, analysis can vary significantly according to the software used to process, render, or analyze it. Review of a data set would likely require an examination of the software code used to process the data as well. All of these criteria create more work for already burdened reviewers. Who is responsible for this review is often an open question. Many publications require deposit of the data into trusted repositories. Some repositories, such as [DRYAD](), [NASA Planetary Data System,]() and the [Qualitative Data Repository](), perform data review upon deposit, but many others do not. Journals increasingly require data submission as part of publication process, and some that require data release with publication require at least cursory peer review of associated data with their papers. Other traditional journals are publishing "data papers." These data papers are "scholarly publication of a searchable metadata document describing a particular on-line accessible dataset, or a group of datasets, published in accordance to the standard academic practices," as defined by [Chavan and Penev]() in 2011. Finally, a number of data journals are focused solely on the sharing of research data sets. Those publications explicitly provide peer review prior to "publication" of the data.

In a [2014 paper, Candela et al. identified 116 journals]() that publish data papers. Seven of those were "pure" data journals, i.e., those that focus only on data papers, and 109 published at least one data paper. Of those 109, Springer/BioMedCentral contributed 94 titles. Candela reviewed a variety of characteristics of those titles, even their peer review criteria, but at a general level. Given the variance and the important implications of specifics in the peer review process, it is important to explore these policies in greater detail. To undertake this study, I began with the Candela et al. list of titles not published by Springer. In the three years since Candela's paper was published, additional titles have launched and more detailed data access policies are being advanced. To the original list, I then added the 13 newly released data journals, so in total 39 peer review policies were reviewed. Beyond this, SpringerNature/BioMedCentral now has a [set of four data policies that cover almost 750 of their titles](#) and that cover a sliding scale of increasing rigor in data sharing. Of those 746 titles, I included only the 8 journals in this study that use the Level 4 policy, which stipulates the highest requirements for data sharing, rather than all 746 titles. For each of these 39 journals, the available peer review criteria or instructions were collected and reviewed. I then extracted key criteria from each to compile and compare similarities and differences.

The peer review polices fell into five broad categories of review criteria for data sets: Editorial, Metadata, Data Quality, Methodology Review, and Other. Nine criteria fell under Editorial review; these focused on the relatedness of the dataset to the journal's scope, importance, and other subjective considerations. Metadata Quality criteria included 9 elements focused on empirical quality. The third major category examined the methodology information provided with the dataset, about how the dataset was created. The final category included 11 miscellaneous criteria that fell outside the first three categories.

**Editorial Review of the Dataset**

Editorial review criteria were evident in nearly every peer review policy, as 36 of 39 policies mentioned an initial editorial check of overall quality. Topical appropriateness for the publication and suitability within the scope of the journal were mentioned in the vast majority of policies, 29/39 and 28/39 times respectively. Novelty of the science described was also an important factor that was mentioned in 27 polices, though one journal interestingly stated specifically that novelty was not a criterion for publication. Several of the tiles also included open peer review either explicitly or as an option if the reviewers wished, and this element was included in the Other category. Seventeen policies explicitly required a conflict of interest statement.

| **Criterion** | **Included in Policy** |
|---|---|
| Editorial Review | 36 |
| Open Peer Review | 3 |
| Conflict of Interest Policy | 17 |
| Topical Appropriateness in Title | 29 |
| Suitability for Publication in Title | 28 |
| Importance of the Subject | 12 |
| Overall Quality of Research | 31 |
| Unpublished | 13 |
| Originality/Novelty of Science | 27 |

**Metadata review of the data set**

After a general editorial review, metadata review of the data set was the most comprehensively inclusive criteria set in most peer review policies, perhaps because it is the easiest to review objectively. Overall metadata quality was mentioned in 33 of 39 policies. The clarity of the writing of any abstract, title or additional materials was the second most important criterion in this category. The presentation of the metadata (24/39), its completeness (19/24), and its conformance either to a journal's template or to community standards (19/24) were the next most important criteria. Despite the number of journals expecting that data be deposited either directly with the journal or in a public repository, the number of policies that required a DOI for the data set (10/39) was considerably lower than one might expect. Rights information was only mentioned in one policy, although this is likely because of the high number of policies that required open licensing, as tracked in the Other category below.

| Criteria | Included in Policy |
| --- | --- |
| Metadata Quality | 33 |
| Metadata Presentation | 24 |
| Metadata Standards Conformance/ Template Conformance | 19 |
| Title/Abstract/Writing Clarity | 25 |
| Keyword Selection | 2 |
| Dataset DOI Assignment | 10 |
| Metadata Rights Information | 1 |
| Provenance | 0 |
| Metadata Completeness | 19 |

**Data review of the dataset**

A review of the data in a data set was mentioned in all but one of the data review policies, but there was not a lot of consistency across the polices as to what practices should be included in the review. The ability to reuse the data (23 of 39 policies) was mentioned most regularly in the context of reviewing the dataset within the data paper review criteria, but equally apply to the methodological section. Taken as a whole, reviewers are asked to consider, based on the sum of the information provided, whether someone else could replicate these results. The reviewers were instructed to review units of measure in the dataset (22/39), and assess whether there were any errors in technique, fact, calculation, interpretation, or completeness (18/39). Thereafter, the overall logical and consistency of the data (17/39), the data format's consistency (17/39), whether basic data quality methods were followed, whether the data selection was appropriate (i.e., whether it was specific or broad enough for the specific use) (16/39), and whether the software used to analyze the data was described in sufficient detail (15/39). This group showed the was the most variety in criteria. Several criteria were only mentioned a few times, such as plausibility (1/39), high quality (2/39), documentation of data anomalies found (2/39), outliers (1/39), checksums verified (3/39), or anonymization processes run (5/39).

| **Criteria** | **Included in Policy** |
|---|---|
| Data Logic & Consistency | 17 |
| Data Format Consistency | 17 |
| Data - Non-Proprietary Formats | 6 |
| Data - Plausibility | 1 |
| Data - of High Quality | 2 |
| Data - Worthy of sub selection/broad enough/Appropriate selection | 16 |
| Data reuse | 23 |
| Data - Software Used to Process Described | 15 |
| Data Units of Measure | 22 |
| Data Quality Methods | 17 |
| Data Anonymization | 5 |
| Data Anomalies Documented | 2 |
| How are outliers identified & treated? | 1 |
| Any data errors introduced by transmissions (checksums)? | 3 |
| Any errors in technique, fact, calculation, interpretation, completeness? | 18 |

**Methodology behind creation of the dataset**

The methodology used to create a dataset is as critical to understanding the resource as the data itself. While there was more consistency about inclusion of methodology review criteria than about metadata, there were more policies that didn't include reference to methodology at all (10/39) than was the case with metadata. The most-referenced criteria regarding methodology were: was the methodology was appropriate for the study being undertaken (25/39)? Were the data collection methods adequately described (23/39)? Could the process be replicated based on the information provided (23/39)? And lastly, did the process conform to high technical standards (18/39)? Interestingly with regard to methodology, few criteria specifically focused on the equipment used (2/39) or on quality-control processes (3/39).

| Criteria | Included in Policy |
|---|---|
| Methodology - Appropriate | 25 |
| Methodology - Current | 9 |
| Methodology - Data Collection Methods | 23 |
| Methodology - High Technical Standard | 18 |
| Methodology - Equipment Description | 2 |
| Methodology - Replicable | 23 |
| Methodology - Quality Control | 3 |
| Study design - Overall | 16 |
| Methodology - Supporting Experiments | 16 |
| Study design - Replications Identified | 15 |
| Study design - Independencies or Co-Dependencies | 16 |
| Study/Data Preprocessing Described | 16 |

**Other Review Criteria**

A very diverse set of criteria didn't fit into the other major groupings of review features, one theme of which was the availability of the data and its openness, whether as part of the reviewing journal or in other environments, such as repositories, and the status of those repositories. The majority of polices (32/39) had some public accessor open license requirement. If the data set wasn't included in the publication, a link to the data source (31/39) and descriptions of how to access the data (30/39) were most often included in the review criteria. The citation to other relevant materials, papers or research was an important review element (29/39), indicating that data papers are expected to have as much citation backing as traditional papers. Experimentation ethics (28/39) was also frequently included as a criterion for review, not only for human subjects but for animal subjects as well. Whereas only 3 of the titles required an open review practice, 15 provided for the anonymity of reviewers where open peer review was an option.

| Criteria | Included in Policy |
|---|---|
| Citations to Other Relevant Materials | 29 |
| Fairness | 1 |
| Anonymity of Reviewers (if desired) | 15 |
| Ethics of Experimentation | 28 |

| Public Data Sharing, Open License Requirement | 32 |
| Data Repository with Sustainability Model | 13 |
| Data Sharing - Platform Agnostic | 1 |
| Link to Public Repository | 31 |
| Descriptions of How to Access Data | 30 |
| Abbreviations Noted/Defined | 15 |
| All Contributors/Authors Credited | 16 |

**Most- and Least-referenced Peer Review Criteria**

Combining all of the criteria, we can then identify those that are included in the most and fewest peer review policies examined. No single criterion was included across every one of the peer review policies. There was diversity among the most-included policies, but a high-level focus on general appropriateness and quality, overall metadata quality, and suitability were broadly referenced. Review of the accessibility and public licensing of the content were also widely included. The ethical processes of data collection was also widely found. Several criteria were mentioned in only one policy. One policy recommended by Lawrence et al, on provenance tracking information, wasn't included in any document seen.

**Most-referenced peer review policies**

| Criteria | Included Statement | Lawrence et al |
|---|---|---|
| Editorial Review | 36 | |
| Metadata Quality | 33 | x |
| Public Data Sharing, Open License Requirement | 33 | |
| Overall Quality | 31 | |
| Link to Public Repository | 31 | |
| Descriptions of How to Access Data | 30 | |
| Topical Appropriateness | 29 | |
| Citations to Other Relevant Materials | 29 | |
| Suitability for Publication | 28 | x |
| Ethics of Experimentation | 28 | |
| Originality/Novelty of Science | 27 | x |
| Title/Abstract/Writing Clarity | 25 | |
| Methodology Appropriateness | 25 | x |
| Metadata Presentation | 24 | x |
| Data Reuse | 23 | |

**Least-referenced Peer Review Criteria**

| Criterion | Included Statement | Lawrence et al |
|---|---|---|
| Any data errors introduced by transmissions (checksums)? | 3 | x |
| Methodology - Quality Control | 3 | |
| Keyword Selection | 2 | |
| Data - Of High Quality | 2 | |
| Data Anomalies Documented | 2 | x |
| Methodology - Equipment Description | 2 | |
| Metadata Rights Information | 1 | x |
| Data - Plausibility | 1 | x |
| How are outliers identified & treated? | 1 | x |
| Fairness | 1 | |
| Data sharing - Platform Agnostic | 1 | |
| Provenance | 0 | x |

**What Constitutes Robust Peer Review?**

Based on this study, one can draw a few conclusions about current practice and what constitutes robust peer review of data. First of all, editorial cohesion is still quite important in journal publication and this is also true of the new data journals. This cohesion helps in some ways with discovery and peer review expertise, but could limit the opportunities for novel analysis by combining data sets. More important than the internal consistency and precise review of the data set is a focus on the openness and availability of the data itself. Opportunities for reuse, links to public repositories, and descriptions of how to access the data are of significant importance. This might be viewed as putting into practice the notion that 'reuse is peer review of data.' If data is of sufficient quality that it will be reused, it passes a post-publication peer review. According to [Parsons et al](), "data use in its own right provides a form of review. If data are broadly used and this use is recorded through citation, it indicates a certain level of confidence in the data." There is also a lot of peer-review attention to the ethical concerns of data collection, especially in domains that use human-subject data. Finally, rather than focusing on the quality of the data itself, peer review is more focused on the overall quality of the metadata. There are domain-specific variations on the details of metadata quality, but peer reviewers across the board are instructed to review metadata quality.

    It is interesting to compare the expectations regarding what peer review of data should entail what it actually involves. Of the 23 criteria outlined by Lawrence et al, only 5 were in the top third of criteria being applied by journal publishers today. Forty-three percent, ten of the 23, were in the middle third of criteria by number of titles applied. Almost a quarter of their criteria, 8 of 23, are in use by fewer than a third of reviewed policies. A significant number of the criteria proposed by Lawrence et al focused on quality measures of the data itself, such as the plausibility of the data, the selection of the data (compared to the universe of the data, its provenance, and the identification of anomalies in the data set. Although there is so much variance between reality and the proposal put forward by Lawrence et al, this should not be seen as a criticism of their work, which has been influential. In some ways, the suggestions put in that paper would be particularly challenging to undertake in practice at scale, as it would

involve work that few reviewers have the time or expertise to perform fully. The variance also shows how much the process of data publishing has changed in the past 6 years.

Another interesting comparison would be to view what journals are doing compared with what researchers would expect to see regarding data publication. In [a 2015 paper by John Ernest Kratz and Carly Strasser](), a survey of researchers described their expectations as to what published research data should include. The most frequently observed feature of a published dataset was open availability (68%), availability in a repository (54%), and the indication of links between the data and a paper (such as via a DOI). Rich metadata (39%), unique identifiers (39%), and formal metadata (25%) were less frequently cited. Interestingly, only 28% of respondents felt that peer review was a core value-add of data publication. On the question of peer review, respondents were most focused on a review of the appropriateness of methods (90%) and metadata to support reproducibility (80%). A deep technical review (75%) and the plausibility of the data (61%) were highly cited. By comparison, the actual peer review process doesn't match the expectations expressed by researchers in this study. While a review of methods was considered most important, this was included in only 64% of policies in any form. Another striking expectation is that the review should include some view on plausibility, but only one policy explicitly stated that as a review criterion. Metadata standards compliance was relatively under-valued, but it was included in 49% of policies. The expectations of peer review by researchers is quite high, but the actual polices seem to be focused more on easily assessable qualities than those that match researcher expectations.

Like every study, this one has its caveats regarding the data collected and the implications of the results. Key among these caveats is that the entire peer review process is not discernable from this study. Not every peer review policy is detailed, and policies could be more or less robust than stated. A lot more robust processes could be masked by vague policies. For example, one journal's catch-all instruction was "that all aspects of the Data Paper and associated dataset will be peer reviewed," which could encompass a great deal. I focused here only on publicly available peer review policies, not on the actual practices of peer reviewers. Also, not every peer review policy is publicly available or detailed. Almost certainly, details provided to peer reviewers about what should be considered in the review are subtly different from the general instructions available online. Also, actual peer reviewer behavior may not be in full compliance with related policies.

In addition, there is robust discussion about the meaning of "dataset validation" in the context of scholarly data sharing. A variety of techniques make it possible to validate the consistency and appropriateness of a data set. Some repositories conduct a data validation process upon deposit, but many do not. Assante et al describe data validation as being "at the moment among the most debated and undefined data publishing phases," and state that "there are no shared established criteria on how to conduct such review and on what data quality is." Sarah Callaghan and her colleagues draw a distinction between the technical and scientific quality components of data quality. They say that technical quality is achieved when data sets come as a package that includes the complete dataset, robust metadata, and appropriate file formats. This is distinguished from scientific quality, which is focused on appropriate collection methods and high overall believability.

Over the past five years, there has been a significant growth in the publication and sharing of scholarly research data, and a concomitant increase in the demand for peer review of data sets. Some organizations and publications are leading by example, and it is valuable to review their efforts to draw from their expertise. Seemingly, those things that are the most difficult to assess are less likely to be included, while those that are easiest to assess are most prevalent. While each domain is different, and the expectations of data review are distinct, as they are with article peer review processes, there are commonalities and baseline standards can be propagated. One hopes that this report provides some basis to develop those baselines.

---

[1] Todd Carpenter is Executive Director of the National Information Standards Organization (NISO). He also serves as the Co-Chair of the Research Data Alliance, RDA/WDS Publishing Data Interest Group.